\documentclass[conference]{IEEEtran}
\usepackage{float}
\IEEEoverridecommandlockouts
\usepackage{amsmath,amssymb,amsfonts}
\usepackage[ruled]{algorithm2e}
\usepackage{algpseudocode}
\usepackage{graphicx}
\usepackage{textcomp}
\usepackage{graphicx}
\usepackage[style=ieee]{biblatex}
\usepackage{booktabs}
\usepackage[table,xcdraw]{xcolor}
\usepackage[normalem]{ulem}
\useunder{\uline}{\ul}{}

\def\BibTeX{{\rm B\kern-.05em{\sc i\kern-.025em b}\kern-.08em
    T\kern-.1667em\lower.7ex\hbox{E}\kern-.125emX}}
\begin{document}

\title{SKALD: Scalable K-Anonymisation for \\ Large Datasets\\

}

\author{\IEEEauthorblockN{Kailash Reddy,
Novoneel Chakraborty,
Amogh Dharmavaram,
Anshoo Tandon}
\IEEEauthorblockA{Email: kailash.nanaluri@gmail.com, cnovoneel@gmail.com, amogh.dharma1@gmail.com, anshoo.tandon@gmail.com}
\IEEEauthorblockA{IUDX/CDPG (FSID), IISc Bengaluru}
}

\maketitle
\begin{abstract}
    Data privacy and anonymisation are critical concerns in today's data-driven society, particularly when handling personal and sensitive user data. Regulatory frameworks worldwide recommend privacy-preserving protocols such as k-anonymisation to de-identify releases of tabular data. Available hardware resources provide an upper bound on the maximum size of dataset that can be processed at a time. Large datasets with sizes exceeding this upper bound must be broken up into smaller data chunks for processing. In these cases, standard k-anonymisation tools such as ARX can only operate on a per-chunk basis. This paper proposes SKALD, a novel algorithm for performing k-anonymisation on large datasets with limited RAM. Our SKALD algorithm offers multi-fold performance improvement over standard k-anonymisation methods by extracting and combining sufficient statistics from each chunk during processing to ensure successful k-anonymisation while providing better utility.
\end{abstract}


\section{Introduction}
In the modern era, data is a pseudo-currency that forms the foundation of digital and social transactions. However, the data economy is fraught with significant privacy risks~\cite{Dwork2017},~\cite{Abowd2023} that threaten individual confidentiality. This makes safeguarding the privacy of data containing private and sensitive data a key issue if the data is to be shared, analysed or released. 

Globally, regulatory pressures drive the data anonymisation imperative, and advisory guidelines are provided on techniques and execution of anonymisation~\cite{Gadotti2024}. In India, the Digital Personal Data Protection (DPDP) Act~\cite{GovernmentofIndia2023} is a precursor to more stringent and clearly outlined legal frameworks around anonymisation. Canada, the EU and the UK stipulate that data should be rendered anonymous such that the data subject is no longer identifiable, while Singapore delineates acceptable privacy parameter ranges for the distribution of anonymised data through well-established privacy-preserving techniques such as k-anonymisation~\cite{Sweeney2002},~\cite{DSCI2023}. 

The General Data Protection Regulation (GDPR), long considered the gold standard for extensive data security measures, outlines stringent guidelines that emphasise the significance of data autonomy~\cite{EuropeanParliament2016a}. The transfer of personal data to third parties or international organisations is strictly regulated. Local processing of data is encouraged to ensure an auditable chain of oversight and responsibility which implies the need for dedicated hardware systems to carry out anonymisation tasks. The available hardware resources provide an upper bound on the maximum size of dataset that can be processed at a time. Large datasets with sizes exceeding this upper bound must be broken up into smaller data chunks for processing. Although optimal k-anonymisation implementation requires knowledge of the entire dataset, standard k-anonymisation tools such as ARX can only operate on a per-chunk basis if the dataset is too large. This lack of a global view of the dataset can negatively affect the utility of the output information. 


In this paper, we extend our work in~\cite{Chakraborty2025} on data de-identification pipelines to propose a novel algorithm - SKALD (Scalable K-Anonymisation for Large Datasets), that operates efficiently when large datasets are broken into smaller chunks due to constraints on the main memory (RAM). We use the loss metrics proposed in~\cite{ElEmam2009} to demonstrate that SKALD offers multi-fold performance improvement over standard k-anonymisation methods, such as the open-source ARX software~\cite{Prasser2016}, by extracting and combining sufficient statistics from each chunk during processing to ensure successful k-anonymisation while providing better utility.

\section{Preliminaries and Prior Work}
\subsection{K-Anonymisation}
K-Anonymisation is a well-established syntactic model of privacy that is popularly used to anonymise releases of tabular data~\cite{Sweeney2002_2, Meurers2021, Garfinkel2023}. To understand how k-anonymisation works, we need to understand the following terms:
\subsubsection{Direct Identifiers}
Direct identifiers are attributes that can be used to directly identify an individual in a dataset.
\subsubsection{Indirect Identifiers or Quasi-Identifiers}
Quasi-identifiers (QIDs) can be used to identify an individual in a dataset when combined with other attributes, but not on their own~\cite{Motwani2007}. A famous example of identification using QID is given by Sweeney~\cite{Sweeney2000} - 87\% of people in the United States were identifiable by a tuple of their $\{$5-digit ZIP, gender, date of birth$\}$ in the year 2000. QIDs are of two types - numerical and categorical.
\subsubsection{Sensitive Attributes}
A sensitive attribute refers to any piece of information that, if disclosed, could lead to privacy violations or harm to an individual. 
\subsubsection{Equivalence Classes}
An equivalence class (EC) in this context is a group of records that share the same values for a set of QIDs.

 In a k-anonymised release, each equivalence class must contain at least k records. This is achieved by generalising QIDs to a less granular representation until the k-anonymity requirement is met, thus reducing the risk of identifying individuals uniquely. 
\subsubsection{Record Suppression}
Record suppression in the context of k-anonymisation is a method used to remove records to improve the utility of a k-anonymisation solution.

 \subsection{The Generalisation Lattice}
 In general, finding an optimal k-anonymity solution is NP-hard~\cite{Meyerson2004}, and previous works have presented solutions involving traversal of a generalisation lattice to find a k-anonymisation solution~\cite{ElEmam2009, LeFevre2005}. A generalisation lattice is a structured representation of all the possible ways to generalise the QIDs of a dataset. Each node in the lattice represents a single transformation that defines a specific resolution for each of the QIDs. The resolution becomes coarser as the generalisation levels increase from bottom to top. An arrow or edge in the lattice indicates that a transformation is a direct generalisation, meaning that the node can be derived by incrementing the generalisation level of a single QID from the predecessor. The bottom-most node of the lattice represents the finest resolution, while the top node represents the coarsest resolution, or the maximally generalised dataset~\cite{Prasser2016}. In general, the information loss increases from the bottom to the top of the lattice. This lattice represents the search space for finding the k-minimal node - the k-anonymous node with the lowest information loss in the lattice.
 
 \begin{figure}[h!]
 \centering
\includegraphics[width=0.4\textwidth]
{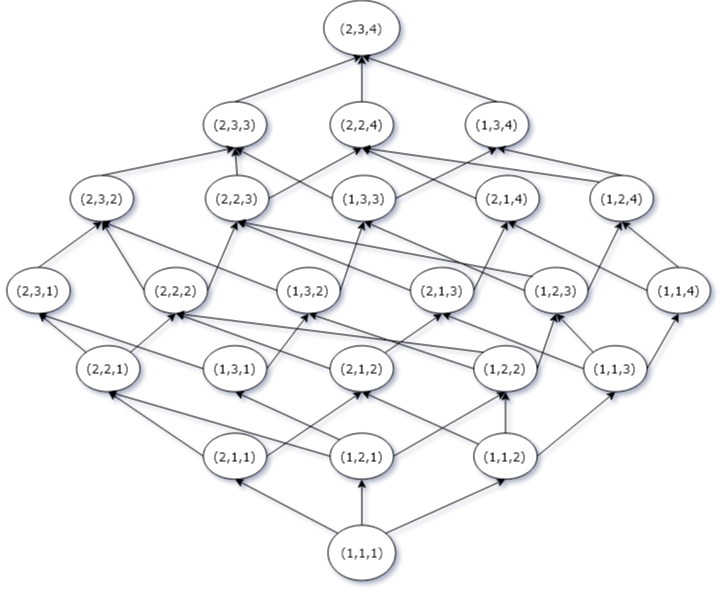}
\caption{An generalisation lattice with 3 QIDs generalised to 2,3 and 4 levels respectively.}
\label{generalisationlattice}
\end{figure}

 The size of this search space can be quantified as the product of the number of levels for each QID's generalisation hierarchy, and grows with the number of QIDs for which these hierarchies are defined. For example in Fig.\ref{generalisationlattice} there are 3 QIDs that are generalised to 2, 3 and 4 levels respectively, and the total number of nodes in the generalisation lattice is 24.
 
\subsection{Prior Work and Their Limitations}
Flash~\cite{Kohlmayer2012}—an existing globally-optimal k-anonymity algorithm—relies on the ability to hold the entire dataset in "main-memory" (referring to the primary-volatile memory type RAM) to build the required memory layouts. Flash~\cite{Kohlmayer2012} mentions that in the case of limited main-memory or very large datasets, a disk-based implementation might be required, which has not yet been developed. A complementary strategy for high-dimensional datasets, Lightning~\cite{Prasser2016}, provides a heuristic, best-effort solution. However, dataset dimensionality is not a direct determinant of dataset volume, and algorithms such as Lightning, Flash, and Incognito~\cite{LeFevre2005}, may still struggle when the dataset size far exceeds the available RAM. These algorithms can perform k-anonymisation only on the portion of the data that fits in the available physical memory. Thus, the solutions obtained are far from optimal. In comparison, we offer a solution that outperforms these algorithms with the same constraints.

\section{SKALD: Our Proposed Solution}
In this paper, we focus on a real-world scenario where the dataset size exceeds the amount of RAM available. The data thus needs to be parsed in chunks to be anonymised. Although each chunk only gives us a limited view of the dataset, we require a global view of all the chunks for efficient k-anonymisation. To solve this problem, we present the SKALD algorithm that is capable of providing a significantly improved solution when presented with batched datasets. Our algorithm is able to parse through chunks of data to retain and combine sufficient statistics from each chunk to k-anonymise the entire dataset without needing to hold the entire dataset in main memory at once. In order to do this, we make use of the fundamentals of the Optimal Lattice Anonymisation (OLA) developed in~\cite{ElEmam2009} for navigation of the generalisation lattice, while modifying it to suit our methodology.

\subsection{Bounding the Number of Histogram Bins}

The amount of available physical memory (RAM) constrains the space available for a histogram of a dataset. Equivalently, this imposes a bound on the maximum number of bins in the histogram, denoted by \(N_{RAM}\). Each bin corresponds to one equivalence class—a unique combination of values across all QIDs.

Consider a dataset \(D\) with \(L\) categorical and \(M\) numerical QIDs. For the \(i\)th categorical QID at resolution \(R_i\), let \(C_i[R_i]\) be the number of unique categorical values. For each numerical QID \(j\), let \(Q_{j,\max}\), \(Q_{j,\min}\), and \(W_j\) denote its maximum value, minimum value, and chosen bin width, respectively.
For these parameters, the total number of bins (equivalence classes) in the histogram, denoted by \(N_{data}\), is given by

\begin{equation}
\label{numEquivalenceClass}
N_{data}
\;=\;
\Bigl(\prod_{i=1}^L C_i[R_i]\Bigr)
\;\times\;
\prod_{j=1}^M \frac{\,Q_{j,\max} - Q_{j,\min} + 1\,}{\,W_j\,}
\;+\; 1,
\end{equation}

\noindent where:
\begin{itemize}[\leftmargin=1.5em]
  \item \(\prod_{i=1}^L C_i[R_i]\) is the number of unique combinations of categorical QID values at resolutions \(\{R_i\}\).
  \item \(\prod_{j=1}^M \bigl((Q_{j,\max} - Q_{j,\min} + 1) / W_j\bigr)\) is the number of numerical‐QID bins at widths \(\{W_j\}\).
  \item The “\(+1\)” term accounts for suppressed records as a separate bin.
\end{itemize}

The parameters \(\{R_i\}\) and \(\{W_j\}\) in \eqref{numEquivalenceClass} are chosen to satisfy
\begin{equation}
\label{numEquivalenceClassShort}
N_{data} \;\le\; N_{RAM}.
\end{equation}

Next, let \(n\) be the number of records per chunk and \(d\) be the size (in bytes) of a single record. Then the chunk occupies
\[
S \;=\; n \times d \quad (\text{bytes of RAM}).
\]
As SKALD operates by extracting and combining sufficient statistics across all chunks, the accuracy of SKALD does not depend on the chunk size. We can ensure that the storage of histograms in SKALD does not require additional memory by operating with smaller chunks. In particular, we can use a reduced chunk size of $n/2$ records occupying $S/2$ bytes, and the remaining $S/2$ bytes can be split between two histograms—i.e.\ each histogram can use at most \(S/4\) bytes. Since each bin in a histogram requires a 4 byte integer to store the count, the maximum number of bins per histogram satisfies

\[
N_{RAM} 
\;=\; 
\frac{S/4}{4}
\;=\;
\frac{S}{16}.
\]

Substituting \(S = n \times d\) yields the RAM‐based upper bound

\begin{equation}
\label{ramBound}
N_{RAM}
\;=\;
\frac{n \times d}{16}.
\end{equation}

Equations \eqref{numEquivalenceClassShort} and \eqref{ramBound} together force

\[
N_{data} 
\;\le\; 
\frac{n \times d}{16}.
\]

Therefore, when choosing categorical resolutions \(\{R_i\}\) and numerical bin widths \(\{W_j\}\), one must retain as much granularity as possible while ensuring
\[
\Bigl(\prod_{i=1}^L C_i[R_i]\Bigr)
\;\times\;
\prod_{j=1}^M \frac{\,Q_{j,\max} - Q_{j,\min} + 1\,}{\,W_j\,}
\;+\; 1
\;\le\;
\frac{n \times d}{16}.
\]

The detailed procedure for selecting \(\{R_i\}\) and \(\{W_j\}\) under this constraint is described in the subsequent sections.

\subsection{The Three Phases of the SKALD Algorithm}
The SKALD algorithm can be broken down into three execution phases explained below.

\begin{figure*}[h!]
\centering
\includegraphics[width=0.7\textwidth]{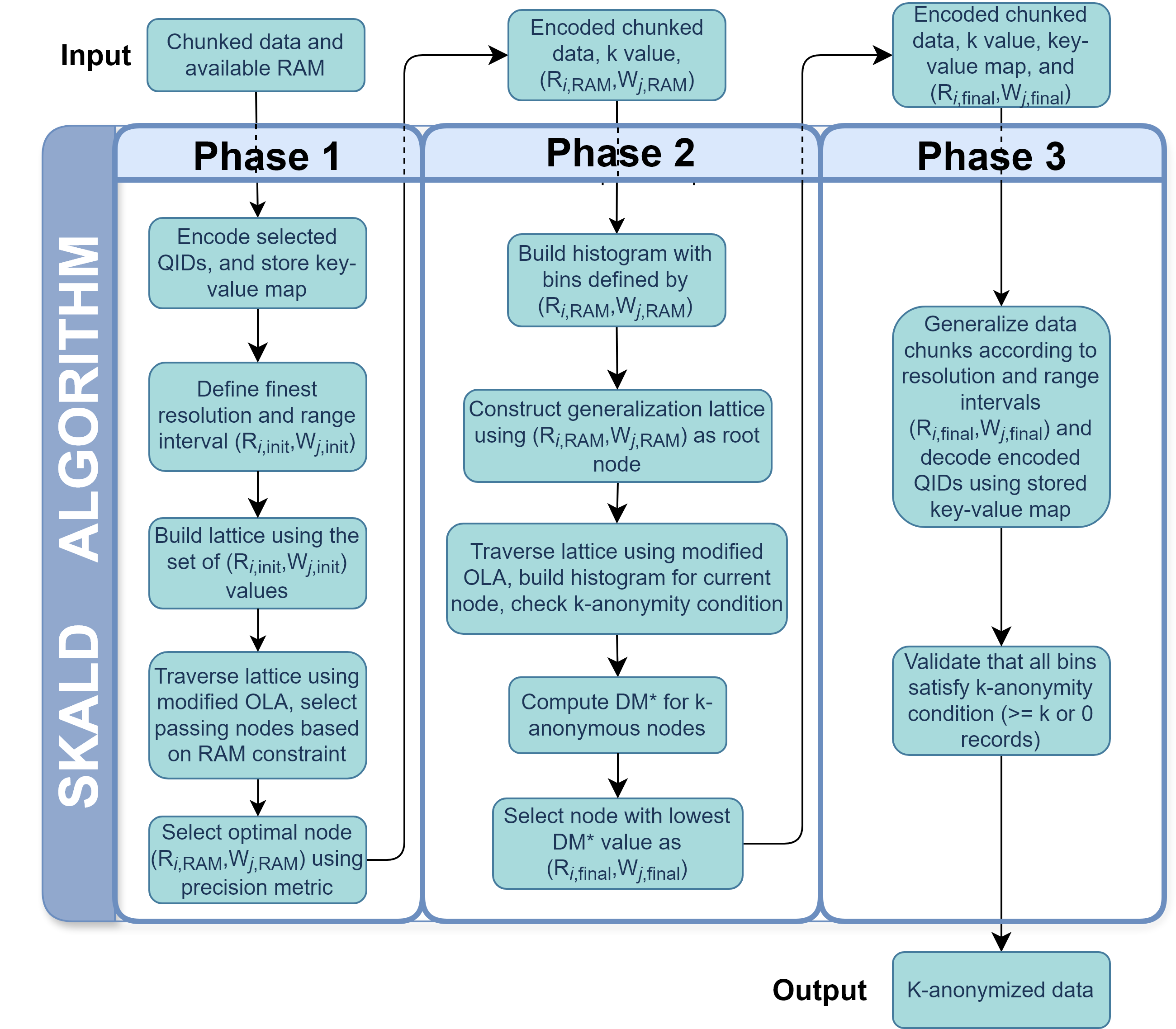}
\caption{The SKALD algorithm}
\label{SKALDalgo}
\end{figure*}

\subsubsection{Phase 1: Encoding Numerical QIDs and Choosing $(R_i, W_j)$}
There are two operations in the first phase. The first is an encoding procedure, wherein sparse numerical QIDs are encoded to reduce the total number of histogram bins. If there is a large discrepancy between the maximum number of possible values and the number of values present in the dataset, then there is a chance that successful k-anonymisation will require a larger bin width than desired for that particular QID. This  increases the amount of information loss in the anonymised dataset and necessitates the use of encoding to reduce the sparsity. To perform this encoding, we pass the chunks of data one at a time, taking the unique values of the numerical QIDs to be encoded in the dataset, sorting the values, and then assigning a new identifier to that value. A key-value map is created during the encoding process and is later used for decoding.

For the second operation, we need to determine appropriate values for $(R_i, W_j)$ to retain as much fine-grained information as possible while satisfying \eqref{numEquivalenceClassShort}. We begin by defining the finest possible resolution $R_{i, \mathrm{init}}$ for categorical QIDs and finest bin width $W_{j, \mathrm{init}}$ for numerical QIDs.

For categorical attributes, a generalisation hierarchy is used to progressively abstract information at different levels. For example, a specific profession such as \textit{Mixed Media Artist} can be generalised to \textit{Artist} at the first generalisation level, \textit{Creative} at the second, and a broader category such as \textit{Non-Technical Sector} at the third. Each level of abstraction reduces the granularity of information, ensuring that categorical attributes can be generalised in a structured manner. For these attributes, $R_{i, \mathrm{init}}$ is chosen as the finest resolution that retains all the categorical value options in the dataset.

For numerical attributes, the finest bin width is determined by the semantic meaning of the unit of measurement. For instance, attributes such as \textit{Age}, typically recorded in whole \textit{years}, or \textit{Weight}, expressed in \textit{kilograms} or \textit{pounds}, naturally have a finest bin width of 1 unit, as further granularity is not meaningful in most analytical contexts. Similarly, for a QID such as \textit{Latitude}, which is measured in \textit{degrees} with up to six decimal places of precision, the finest bin width $W_{j, \mathrm{init}}$ would correspond to $10^{-6}$ degrees, aligning with the smallest meaningful distinction in spatial coordinates.

Using this $(R_{i, \mathrm{init}}, W_{j, \mathrm{init}} )$ pair, we can construct a generalisation lattice where the amount of generalisation increases with every level. The lattice is then traversed, starting from the middle node of the middle level. We select this node as it allows us to eliminate the maximum possible nodes using predictive tagging~\cite{ElEmam2009}. Having labelled all nodes in the lattice as compliant or non-compliant based on \eqref{numEquivalenceClassShort}, we then utilise the precision metric defined in~\cite{Sweeney2002_2} to select the node with the best value of precision among all the passing nodes, labelling this node $(R_{i, \mathrm{RAM}}, W_{j, \mathrm{RAM}} )$. In case two nodes have the same value of precision, we can utilise a pre-determined hierarchy of semantic importance for the QIDs to break the tie.

\subsubsection{Phase 2: Histogram and Lattice Construction}

In the second phase, the data is passed chunk-wise through the algorithm for a second time. A histogram is created to store the counts of users that fall into the bins defined by $(R_{i,\mathrm{RAM}}, W_{j,\mathrm{RAM}})$. This histogram is updated for each chunk, so the final histogram created after processing all the chunks contains aggregate counts over the entire dataset.

A new generalisation lattice is then constructed, using $(R_{i,\mathrm{RAM}}, W_{j,\mathrm{RAM}})$ as the root node. Note that we have the knowledge of the histogram for the root node, i.e.\ $(R_{i,\mathrm{RAM}}, W_{j,\mathrm{RAM}})$, containing aggregate counts over the entire dataset. To apply predictive tagging, we proceed by checking the $k$-anonymity condition from the middle node of the middle level in the generalisation lattice. This requires merging some bins in the root histogram in order to construct the histogram corresponding to the chosen node. We maintain an unmodified copy of the root histogram, and as per the node’s generalisation level we merge the bins in the root histogram to create a new histogram. We now evaluate whether the new histogram satisfies the $k$-anonymity condition. After evaluating a node and marking it as ``pass'' or ``fail'', we proceed to evaluate the next node in the generalisation lattice.

 For all nodes that pass the $k$-anonymity condition within the desired suppression limit, we compute the monotonic Discernibility Metric (DM*) proposed in~\cite{ElEmam2009}. This enables us to select the node with the lowest (best) DM* value, corresponding to minimal information loss, among all the nodes that satisfy $k$-anonymity. We label this node $(R_{i,\mathrm{final}}, W_{j,\mathrm{final}})$.


\subsubsection{Phase 3: Decoding and Generalising the Dataset}
In the third phase the data chunks are passed a final time. In this pass, the encoded QIDs are decoded and the chunks are generalised according to the bin widths defined in $(R_{i, \mathrm{final}}, W_{j, \mathrm{final}}) $. This gives us the final k-anonymised dataset.

The operations within each of the three phases of the SKALD algorithm are compactly summarized in Fig.~\ref{SKALDalgo}.

\section{Experimental Results}
\subsection{Synthetic Data}
We create a synthetic dataset generated using a well-known library for creating pseudo-random but realistic data - \textit{pyfaker}. This dataset is modelled on a real-world medical dataset with attributes that one would expect to find in a sample of data collected from patients by a medical centre or hospital. A snippet of this dataset is shown in Table \ref{tab:sample_synthetic_data}. In this format, we generate 125 chunks of data, each chunk with 1 million rows, where each row corresponds to a medical record of a unique user. This amounts to a total of 125 million records in the entire dataset.
The direct identifiers in this dataset are the ``Patient ID", ``Name" and ``Address" attributes, which can be used to identify an individual without any external information, and are therefore completely suppressed in the anonymised dataset. The dataset has 2 categorical QIDs: (``Blood Group", ``Profession"), 3 numerical QIDs: (``Age",  ``BMI", ``PIN Code"), and the sensitive attribute is ``Health Condition". If a hospital were to publicly release this dataset for analysis, then the attribute being released would most likely be the ``Health Condition", and it would be important to anonymise the dataset to ensure that no individual could be tied to a particular health condition in order to preserve user privacy.

\begin{table*}[t]
\centering
\caption{A sample of the synthetically generated dataset.}
\label{tab:sample_synthetic_data}
\resizebox{\textwidth}{!}{%
\begin{tabular}{|c|c|c|c|c|c|c|c|c|c|}
\hline
\textbf{Patient ID} & \textbf{Name}    & \textbf{Address}    & \textbf{Blood Group} & \textbf{Profession} & \textbf{Age} & \textbf{BMI}     & \textbf{PIN Code} & \textbf{Health Condition}  \\ \hline
P0001-9112-1        & Ekaraj Parikh       & 156, Kala Road                           & O+             & Data Scientist   & 33             & 23.8                & 560044            & Dementia              \\ \hline
P0002-8603-2        & Jack Shankar         & 15/402, Ganesh Chowk                    & A-      & Project Manager & 38                 & 25.3          & 560008            & Dementia                      \\ \hline
P0003-6406-1        & Nakul Prabhakar        & 91/335, Sule Circle                   & B-     & Software Engineer  & 60                 & 25.9          & 560018            & Gout                                 \\ \hline
P0004-8305-3        & Chandani Lanka         & 104, Bahl Marg                       & B-     & Product Manager   & 41        & 23.9               & 561164            & Asthma                       \\ \hline
P0005-9512-2        & Alka Kumar             & 052, Nadig Zila                       & AB-    & Business Analyst   & 29        & 22.2               & 560059            & Asthma                      \\ \hline
\end{tabular}%
}
\end{table*}

\begin{table*}[t]
\centering
\caption{A sample of a 50-anonymised dataset with (2,1),(1, 24, 32)}
\label{tab:k-anonymised_data}
\resizebox{0.75\textwidth}{!}{%
\begin{tabular}{|c|c|c|c|c|c|}
\hline
\textbf{Blood Group} & \textbf{Profession} & \textbf{Age}  &  \textbf{BMI} &     \textbf{PIN Code} & \textbf{Health Condition}  \\ \hline
                           O      & Data Scientist    &    {[33 - 33]}          & {[12 - 36)}        & [560032 - 560063]            &Dementia                      \\ \hline
                           A       & Project Manager   &      {[38 - 38]}      & {[12 - 36)}  & [560000 - 560031]            & Dementia                      \\ \hline
                          B        & Software Engineer  &   {[60 - 60]}         & {[12 - 36)}  & [560000 - 560031]            & Gout                                 \\ \hline
                            B        & Product Manager   &    {[41 - 41]}        & {[12 - 36)}        & [561164 - 561195]            & Asthma                      \\ \hline
                         AB          & Business Analyst   &  {[29 - 29]}      & {[12 - 36)}      & [560032 - 560063]            & Asthma                      \\ \hline
\end{tabular}%
}
\end{table*}
\subsection{Generalisation Hierarchy}
We construct generalisation hierarchies for each QID. These hierarchies define how raw data values can be systematically transformed into more generalised forms across multiple levels. The generalisation levels for both categorical and numerical attributes form the basis for building the anonymisation lattice used in our algorithm. The section below shows how the generalisation hierarchies are built for each QID in our synthetic dataset.

\subsubsection{Categorical Data}
\begin{itemize}
\item \textbf{Blood Group:}  
    \begin{itemize}
    \item Level 1 (Resolution = 1): 
    
        Original blood groups (8 categories): 
    
        \texttt{A+, A-, B+, B-, AB+, AB-, O+, O-}
    
    \item Level 2 (Resolution = 2): 
    
        Generalised by type (4 categories): 
        
        \texttt{A, B, AB, O}
    
    \item Level 3 (Resolution = 3): 
    
        Maximally generalised (1 category):         
        
        \texttt{*} (representing a suppressed value) 
    \end{itemize}

\item \textbf{Profession:}  
    \begin{itemize}
    \item Level 1 (Resolution = 1)
    
        Original professions (16 categories):  
    
        \texttt{Medical Specialists, Allied Health, Nursing, Healthcare Support,\ldots}
    \item Level 2 (Resolution = 2) 
    
        Generalised into broader domains (4 categories): 
    
        \texttt{Healthcare, Education, Creative, Engineering }
    \item Level 3 (Resolution = 3) 
    
        Grouped by sector (2 categories): 
        
        \texttt{Service Sector, Non-Service}
    \item Level 4 (Resolution = 4) 
    
        Maximally generalised (1 category): 
        
        \texttt{*} (representing a suppressed value)
    \end{itemize}

\end{itemize}

\subsubsection{Numerical Data}
\begin{itemize}
\item \textbf{Age:}  
    \begin{itemize}
    \item Level 1 (Bin width = 1): Raw values 
    
        \texttt{33 → [33 - 33]}
    \item Level 2 (Bin width = 2): 
    
        \texttt{33 → [32 - 33]}
    \item Level 3 (Bin width = 4): 
    
        \texttt{33 → [32 - 35]}
    
    \ldots and so on. The bin width doubles at each subsequent level until level 7. The final level which spans the entire domain is the maximally generalised level:
    \item Level 8 (Bin width = 67): 
    
        \texttt{33 → [19 - 85]}
    
    \end{itemize}

\item \textbf{BMI:}
    \begin{itemize}
    \item Level 1 (Bin width = 0.1): Raw values with original precision 
    
        \texttt{22.5, 23.8, 27.6}
    \item Level 2 (Bin width = 1): Values are rounded or grouped into bins of width 1  
        
        \texttt{23.8 → [23 - 24)}
    \item Level 3 (Bin width = 2): 
    
        \texttt{23.8 → [23 - 25)}
    
    \ldots and so on. The bin width doubles at each subsequent level until level 5. The final level which spans the entire domain is the maximally generalised level:
    \item Level 6 (Bin width = 24): 
    
        \texttt{23.8 → [12 - 36)}
    \end{itemize}

\item \textbf{PIN Code:}  
    The synthetic dataset has 1347 unique PIN Codes. Due to sparsity, PIN Codes are first encoded into integers, where the smallest numerical value of the PIN Code in the dataset is mapped to 1 and the largest is mapped to 1347.
    \begin{itemize}
    \item Level 1 (Bin width = 1): 
        
        \texttt{570025 (encoded to 221) → [221 - 221]}
    \item Level 2 (Bin width = 2): 
    
        \texttt{221 → [221 - 222]}
    \item Level 3 (Bin width = 4): 
    
        \texttt{221 → [221 - 224]} 
        
        \ldots and so on. The bin width doubles at each subsequent level until level 11. The final level which spans the entire domain is the maximally generalised level:
    \item Level 12 (Bin width = 1347): 
    
        \texttt{221 → [0 - 1346]} 
    \end{itemize}

\end{itemize}
For example, consider a node in the generalisation lattice, where the $(R_i, W_j)$ pair is represented as ((2, 1),(1, 24, 32)). The first tuple (2, 1) corresponds to resolution of the categorical QIDs, indicating that Blood Group is generalised to Level 2 and Profession to Level 1. The second tuple (1, 24, 32) corresponds to the numerical QIDs, implying that Age is generalised with a bin width of 1, BMI with a bin width of 24, and encoded PIN Code with a bin width of 32.

\section{Results}
To conduct our experiments we use a commodity-class server machine, equipped with a 16-thread (8-core, 16-thread) Intel Core i7-10700 CPU @ 2.90 GHz (up to 4.8 GHz), with a 16 MB L3 cache and 64 GB RAM.

To evaluate the effectiveness of  SKALD, we compare it against ARX, a widely adopted k-anonymisation framework. ARX supports both the globally optimal Flash algorithm~\cite{Kohlmayer2012} as well as the heuristic-based Lightning algorithm~\cite{Prasser2016}. Lightning is tailored for high-dimensional data and trades optimality for speed, making it unsuitable for our evaluation, which prioritizes data utility.  Since our primary goal is to minimize information loss under memory constraints rather than optimize for runtime, we restrict the comparison to Flash.

A representative snippet of the output generated by SKALD for \(k\) = 50 using 125 chunks is shown in Table~\ref{tab:k-anonymised_data}. This configuration resulted in the generalization node [$(R_{i, \mathrm{final}}, W_{j, \mathrm{final}} )$]:  ((2, 1), (1, 24, 32)).

\begin{table}[h!]
\centering
\caption{Results for $k = 50$ across varying no. of chunks}
\resizebox{\columnwidth}{!}{%
\label{tab:chunking_k50}
\begin{tabular}{|c|c|c|}
\hline
\textbf{No. of Chunks} & \textbf{SKALD Generalisation Node} & \textbf{ARX Generalisation Node}\\ \hline
5   & \(((3,1), (1,24,128))\) & \(((3,4), (4,24,4))\)   \\ \hline
25  & \(((3,1), (1,24,32))\) & \(((3,4), (4,24,4))\) \\ \hline
125 & \(((2,1), (1,24,32))\)  & \(((3,4), (4,24,4))\)\\ \hline
\end{tabular}%
}
\end{table}

Table~\ref{tab:chunking_k50} compares the generalisation nodes of SKALD and ARX for different numbers of chunks when \(k = 50\). SKALD algorithm extracts and combines sufficient statistics from each chunk to provide reduced bin widths with an increase in the number of chunks, where each data chunk contains 1 million records. This reduced bin width retains more fine-grained information. ARX, on the other hand, anonymises the data on a per-chunk basis and shows no improvement in the generalisation node with an increasing number of chunks. 

To quantify information loss in the anonymised outputs, we use the \textit{monotonic variant of the Discernibility Metric}, denoted as DM*, and first proposed in~\cite{ElEmam2009}. The metric is defined as

\begin{equation}
\text{DM}^* = \sum_{E \in \mathcal{E}} |E|^2 + s^2
\end{equation}

where,
\begin{itemize}
    \item \(\mathcal{E}\) is the set of equivalence classes,
    \item \(|E|\) is the size of equivalence class \(E\),
    \item \(s\) is the number of suppressed records.
\end{itemize}

A lower value of DM* implies lower information loss. 
To facilitate comparison, we define the $\text{DM}^*$ Ratio as

\begin{equation}
    \text{DM}^* \text{ Ratio} = \frac{\text{DM}^*_{\text{ARX}}}{\text{DM}^*_{\text{SKALD}}}
\end{equation}

A DM* ratio greater than 1 indicates that SKALD achieved lower information loss than ARX for the same parameters.

\begin{figure}[!htbp]
    \centering
    \includegraphics[width=0.5\textwidth]{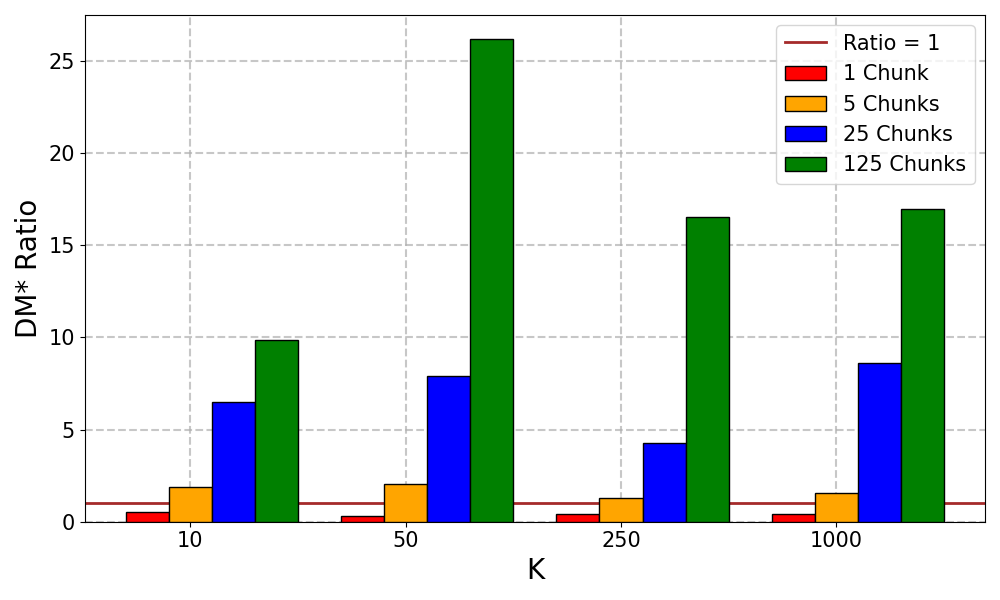}
    \caption{DM* Ratio vs. \textit{k} for different numbers of chunks}
    \label{fig:dm_ratio_k_chunks}
\end{figure}

For our experiments, we varied the anonymity parameter $k \in (10, 50, 250, 1000)$ with a varying number of chunks belonging to the set $\{1, 5, 25,125\}$. Fig.~\ref{fig:dm_ratio_k_chunks} presents the results as a bar chart, and visualises the improved performance of SKALD against ARX with an increase in the number of chunks. In particular, the performance improvement using SKALD for 25 chunks is at least \emph{four-fold} and for 125 chunks is at least \emph{nine-fold} across different values of k. We can infer from these results that the performance gap between SKALD and ARX in terms of DM* increases as the number of chunks and thus, the dataset size, increases.

\section{Conclusions}
Local hardware systems are preferred to carry out anonymisation tasks to ensure an auditable chain of control and accountability. The available hardware resources provide an upper bound on the maximum size of data that can be processed at a time. Large datasets thus need to be broken up into smaller chunks.
In this work we proposed SKALD, a chunk-based k-anonymisation algorithm that is performant within the memory bounds imposed by local hardware. SKALD significantly outperforms ARX's Flash algorithm (in terms of reducing information loss) with an increase in the number of data chunks. SKALD extracts and combines sufficient statistics from each chunk to maintain a global data perspective to provide better output utility. The SKALD algorithm's performance is independent of the size of the data chunks. To perform the comparison, we generated a synthetic dataset modelled on real-world data. Compared to ARX, the results of the experiments on this dataset showed at least a \emph{nine-fold} reduction in the monotonic Discernibility Metric (DM*) value of SKALD for 125 chunks, across different values of k - where each data chunk was designed to contain 1 million records.

Future work includes extending SKALD to support \textit{l}-diversity and \textit{t}-closeness, as well as optimising the computational requirements of the SKALD algorithm.






\printbibliography

\end{document}